\documentclass[12pt]{article}
\usepackage{latexsym,epsfig,graphics}
\newcommand{\be}{\begin{equation}}
\newcommand{\ee}{\end{equation}}
\newcommand{\bq}{\begin{eqnarray}}
\newcommand{\eq}{\end{eqnarray}}
\newcommand{\intst}{\int_{0}^{p^{+}}\! d\sigma \int d\tau}
\newcommand{\st}{\sigma,\tau}
\newcommand{\ust}{u\sigma,u\tau}
\begin{document}
\begin{titlepage}
\today          \hfill 
\begin{center}
\hfill    LBNL-62220 \\

\vskip .5in

{\large \bf Field Theory On The World Sheet: Mean Field Expansion
And Cutoff Dependence}
\footnote{This work was supported in part
 by the Director, Office of Science,
 Office of High Energy and Nuclear Physics, 
 of the U.S. Department of Energy under Contract 
DE-AC02-05CH11231.}
\vskip .50in


\vskip .5in
Korkut Bardakci\footnote{e-mail: kbardakci@lbl.gov}
\vskip 9pt
{\em Department of Physics\\
University of California at Berkeley\\
   and\\
 Theoretical Physics Group\\
    Lawrence Berkeley National Laboratory\\
      University of California\\
    Berkeley, California 94720}
\end{center}

\vskip .5in

\begin{abstract}
Continuing earlier work, we apply the mean field method to the world
sheet representation of a simple field theory. In particular, we study
 the higher order terms in the mean field expansion, and show that 
their cutoff dependence can be absorbed into a running coupling constant.
The coupling constant runs towards zero in the infrared, and the model tends
towards a free string. One cannot fully reach this limit because of infrared
problems, however, one can still apply the mean field method to the
high energy limit (high mass states) of the string. 
\end{abstract}
\end{titlepage}

\newpage
\renewcommand{\thepage}{\arabic{page}}
\setcounter{page}{1}
\noindent{\bf 1. Introduction}

\vskip 9pt

The present work is the continuation of a program developed in a
series of earlier papers [1-7]. The main goal of this program was
to study the sum of planar graphs of a given field theory by first
reformulating it as a two dimensional local theory on the world sheet.
This reformulation makes it possible to do dynamical calculations
using the mean field approximation. With the help of 
 this approximation, it was possible to show string formation
in the case of simplest of field theories, a scalar theory with a cubic
 interaction, thus building a bridge between field theory
and the string theory.

Although we believe that a promising start has been made in addressing
an old and important problem [8], there are still many questions left
to be answered. To name a few, the world sheet parametrization was based
on light cone kinematics, with the resulting loss of manifest Lorentz
covariance. The meanfield method has so far not been applied to more 
realistic theories than the scalar cubic theory. Also, since this method can be
formulated as an expansion in inverse powers of the transverse space
dimensions, the expansion parameter may not be all that small and
its convergence is questionable.

In this article, we will address another problem, which is as important
as any listed above. In doing so, we will also indirectly shed
light on the problem of convergence of the mean field expansion. Both
in constructing the world sheet theory, and also in applying the mean field
method, it was  necessary to introduce  cutoffs. These were
taken to be the spacings of the two grids set up on the world sheet, or
alternatively the corresponding cutoffs in momentum space. Of course,
ultimately one would like to get rid of them by renormalizing the model [9].
However, in this case, renormalization is not a completely straightforward
matter. In particular, one is rather limited in the choice of counter
terms in the initial action. Nevertheless, it was shown in references [4-7]
that, by assigning a  natural cutoff dependence to the original
parameters of the model, in the leading order of the mean field expansion,
 the slope of the emerging string came out
finite. One can then hope that other physical quantities
will also turn out to be cutoff independent.

As is well known, the cutoff dependence of a field theory is
intimately related to its scaling properties. The renormalization
group equations are in general based on this connection. In the
world sheet theory we are studying, there is also a connection
between scaling and Lorentz invariance. This is because the world
sheet parametrization we are using is based on the light cone
variables, with the resulting loss of manifest Lorentz covariance.
In this picture, the boost along the special light cone direction is
implemented by the scaling of the world sheet coordinates. It then
follows that scale invariance is a necessary (but not sufficient)
condition for Lorentz invariance. We therefore see that the questions
of cutoff independence and Lorentz invariance are connected; the
necessary (and for cutoff independence, sufficient) condition for
both is scale invariance.

This paper is mainly devoted to the study of scaling behaviour
and the cutoff dependence of
the world sheet model  that represents the massless scalar with cubic
interaction. The tool we will use will be the mean field expansion,
which is an expansion  in inverse powers
of $D$, the dimension of transverse space. In order to make this paper
self contained,
 in section 2, we will briefly
review the derivation of the world sheet model corresponding to
the massless scalar  field theory with cubic interaction, and
in section 3, we will discuss the
application, in the  leading order, of the mean field method to the
model. Within this approximation, we will be able to demonstrate string
formation on the world sheet. We will also discuss how
the two cutoffs are introduced on the world sheet, and how various fields
and parameters transform under scaling, and how this is related to their
cutoff dependence. The material covered in these
 two sections is mostly a review of the earlier results [1-6].

A crucial step in the derivation of the results reviewed here
 was the minimizing of the ground state energy. Since the
ground state energy is cutoff dependent, one may wonder how much a
result such as string formation depends on the regulation scheme
used. In section 4, we generalize the sharp cutoffs used so far
to a smooth cutoff with an two arbitrary profile functions. We show that
 so long as the profile functions are positive, the result about string
formation remains valid. This section, as well as the rest of the
article, contains  material that is mostly new.

The previous sections only dealt with the leading order in the
expansion in powers of $1/D$. The higher order terms in this
expansion can be computed in two steps: In the first step, we compute
an effective action by 
integrating over the field ${\bf q}$, the world sheet  field that represents
the transverse momenta. In section 5, we show that this effective action
can be written as a generalized sigma model in a new dynamical field $\phi$.
We then expand this action simultaneously
in powers of $1/D$ and in the number of derivatives with respect to the
world sheet coordinates, and we determine explicitly
 the leading term of this expansion,
which provides the potential energy for $\phi$, as well as the next term,
which provides the kinetic energy. These terms were already derived in references
[5,6]. The field $\phi$ plays an important role in the model; for example,
 a non-vanishing
ground state expectation value for $\phi$ generates a non-vanishing
string slope.
  We also investigate the cutoff dependence of the action,
 and  show that, up to logarithmic corrections, the dependence on the powers of
 cutoff is what is expected on the basis of simple power counting.

The next step of the program is to use this sigma model as the starting point
 for computing
higher order contributions.  Here, however, we face the unusual situation
of having an initial action that is cutoff dependent. 
Of course, higher order contributions will generate further cutoff dependence.
This brings up the question, taken up in section 6,
 of how to renormalize the model, so that observables become cutoff
independent. In standard field
theory, divergences are normally cancelled by introducing counter
terms in the original action. Here, since $\phi$ started life as an
auxilliary field in the original action, we do not have the freedom
to introduce arbitrary counter terms.  Before trying to answer this
question, we can ask even a more basic one: What are the observables?
It is natural to assume that in a theory in flat space, all
observable quantities can ultimately be expressed in terms of Lorentz 
invariant ones. In a covariantly formulated theory, this is a
trivial requirement, but here, the situation is different, since it
is not at all straightforward to identify Lorentz invariant objects.
Since Lorentz invariance implies scale invariance, we impose on the
observables the more modest requirement of scale invariance.  Again, by
simple power counting, scale invariant objects are shown
to have no power dependence on the cutoff. Consequently, Lorentz
invariance means no dependence on powers of the cutoff.

However, there still remains a logarithmic dependence on the cutoff,
 to which we turn next. By means of a simultaneous
scaling of both the coordinates and the field $\phi$, we transform the
action into a new form, which no longer has power dependence on the
cutoff. In the new action, the original expansion parameter is replaced
by $\lambda^{2}$, which is inversely proportional to the log of the
cutoff. Also, terms with more than two derivatives are suppressed by
inverse powers of the same log of the cutoff. We note
that, depending on the cutoff, $\lambda^{2}$ could be smaller than
$1/D$ and thus provide a better parameter of expansion.

 In section 7, we consider
the perturbation expansion based on the action above and on an expansion in
 powers of $\lambda$. We show that the higher order contributions
do not change the cutoff structure of the initial action in any
substantial way. The logarithmic dependence of both $\lambda$ and of the
terms with more than two derivatives remain unmodified.

We conclude section 7 with a discussion of these results from the point
of view of the Wilsonian renormalization scheme. Starting with an initial
cutoff $\Lambda$, we try to reduce it to a much lower value $\mu$, by
continually integrating over the high end of the spectrum. As we do this,
the coupling constant, which is inversely proportional to $\ln(\Lambda/\mu)$,
runs towards zero, and the model tends towards  a free
string. As expected, this limit, the infrared fixed point
of the model, is scale invariant
and cutoff independent.  However, this result rests on various
approximations, which are valid only in the ultraviolet regime. 
Consequently, it is not possible to go all the way to the infrared and
reach this fixed point. We argue, however, that by keeping $\mu$ large
enough, we can probe the asymptotic spectrum of the string, but not the
lower lying states. The last section summarizes our conclusions.

\vskip 9pt
\noindent{\bf 2. The World Sheet Action}
\vskip 9pt

This  and the next section are
  devoted to a review of the results obtained in Refs.[1-6]. The
 basic idea is first to represent the sum of planar
graphs of a given field theory as a local field theory on the world sheet,
and then use the mean field approximation to show string formation. 
Starting with a world sheet parametrized by the light cone variables
$$
\tau = x^{+} =(x^{0}+x^{1})/\sqrt{2},\;\; \sigma= p^{+}= (p^{0}+p^{1})/
\sqrt{2},
$$
a general planar Feynman graph for the massless scalar theory with
cubic interaction
in $D+2$ dimensions can be
 represented by a bunch of solid lines (Fig.1)[10].
\begin{figure}[b]
\centerline{\epsfig{file=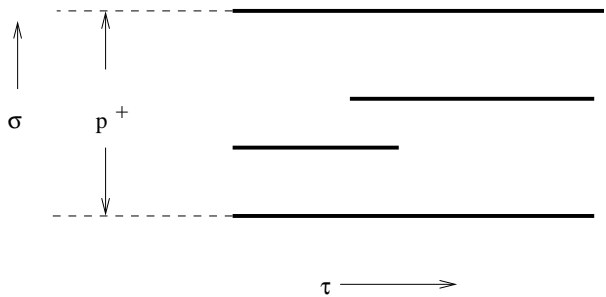,width=8cm}}
\caption{A Typical Graph}
\end{figure}
 One associates a particular
transverse momentum ${\bf q}_{n}$ with the
n'th line, and two adjacent lines $n$ and $n+1$, with
momenta ${\bf q}_{n}$ and ${\bf q}_{n+1}$, represent the light cone
propagator
\be
\Delta(p)=\frac{\theta(\tau)}{2 p^{+}}\,\exp\left(-i\tau\,\frac{{\bf p}^{2}}
{2 p^{+}}\right),
\ee
where ${\bf p}={\bf q}_{n}-{\bf q}_{n+1}$. A factor of $g$ is inserted at the 
 beginning and at the end of each line, where the  interaction
takes place.

The light cone Feynman rules sketched above can be reproduced by a local
field theory on the world sheet. Here, we summarize the  results, and 
 refer the reader to [1-6] for the detailed derivations. The transverse
momenta ${\bf q}$, originally defined only on the solid lines, can be promoted
to a local field ${\bf q}(\st)$ over the whole world sheet. If we think of
the solid lines as boundaries on the world sheet, and the rest as the bulk,
${\bf q(\st})$ satisfies the equation
\be
\partial_{\sigma}^{2}{\bf q}(\st)=0,
\ee
in the bulk, and the Dirichlet boundary condition
\be
\partial_{\tau}{\bf q}(\st)=0
\ee
on the solid lines, since the ${\bf q}$ flowing through a
solid line is constant ($\tau$ independent). With the help of
 a Lagrange multiplier ${\bf y}(\st)$, both the equations of motion and the
boundary conditions are incorporated into the following action:
\be
S_{q}
=\intst\left(-\frac{1}{2} {\bf q}'^{2}+\rho {\bf y}\cdot\dot{{\bf q}}\right),
\ee
where a dot represents the derivative with respect to $\tau$ and a prime
the derivative with respect to $\sigma$. The function $\rho$ is defined to be
the sum of Dirac delta functions with support on the solid lines.

In this expression for $S_{q}$, one has to integrate functionally not only 
over ${\bf q}$ and ${\bf y}$, but also over the positions and the
lengths of the solid lines. This is best accomplished by introducing a
two component fermion $\psi_{i}(\st)$, $i=1,2$, and its adjoint
$\bar{\psi}_{i}$, and setting
\be
\rho=\frac{1}{2}\bar{\psi}(1-\sigma_{3})\psi.
\ee
The action for the fermions is
\be
S_{f}=\intst (i\bar{\psi}\dot{\psi}-D\,g\bar{\psi}\sigma_{1}\psi).
\ee
To see clearly how this works and to avoid singular expressions, it is
best to discretize the sigma coordinate into segments of length $a$. This
discretization is pictured in Fig.2 as a collection of parallel line
segments, some solid and some dotted, spaced distance $a$ apart.
\begin{figure}[t]
\centerline{\epsfig{file=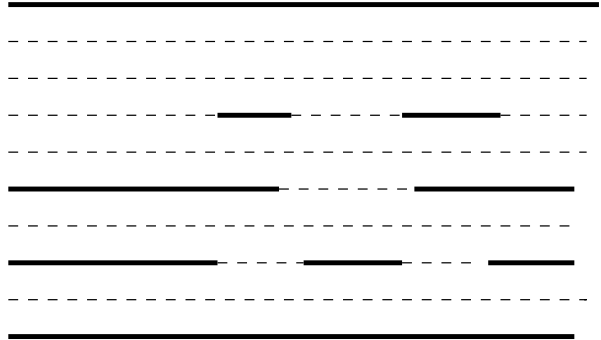,width=8cm}}
\caption{Solid And Dotted Lines}
\end{figure}
 The boundaries are marked by the solid lines, associated with the $i=2$ 
component of the fermion, and the bulk is filled
by the dotted lines, associated with the $i=1$ component. The first term in
Eq.(6) represents the free propagation of the fermion, tracing a solid or a
 dotted line, and the second term, which converts the dotted line into a 
solid line or vice versa, represents the interaction. We have scaled
 the coupling constant $g$ by the dimension of the transverse
space $D$ in anticipation of the large $D$ limit, discussed
in the next section. Finally, integrating
over the fermion field is then the same as summing over the location and the
 length of the boundaries.

There is one more ingredient needed to complete the world sheet action.
In Eq.(4), the part of the integral over ${\bf y}$ that has support in the
bulk (dotted lines) diverges, since the integrand is ${\bf y}$ independent
in this region. To avoid this problem, we add a Gaussian term to action
which cuts off the divergence:
\be
S_{g.f}=\intst\left(-\frac{1}{2}\alpha^{2}\bar{\rho} {\bf y}^{2}\right).
\ee
Here $\alpha$ is a fixed parameter, of which we will say more later on,
and $\bar{\rho}$ is defined by
\be
\bar{\rho}=\frac{1}{2}\bar{\psi}(1+\sigma_{3})\psi,
\ee
and it is complementary to $\rho$: It vanishes on the solid lines and has
support only in the bulk. In fact, on the world sheet regulated by the
grid, one can set
\be
\rho+\bar{\rho}=\bar{\psi}\psi\rightarrow 1/a.
\ee
Finally, the complete world sheet action is given by
\bq
S&=&S_{q}+S_{f}+S_{g.f}\nonumber\\
&=&\intst\left(-\frac{1}{2}{\bf q}'^{2}+\rho {\bf y}\cdot\dot{{\bf q}}
-\frac{1}{2}\alpha^{2}\bar{\rho}{\bf y}^{2}+i\bar{\psi}\dot{\psi}
-D g\bar{\psi}\sigma_{1}\psi\right).
\eq

An important symmetry any light cone action must satisfy follows from
invariance under the boost along the special direction we have labeled
as $1$. This boost translates into the scaling of the world sheet
coordinates by
$$
x^{+}\rightarrow u x^{+},\;\;p^{+}\rightarrow u p^{+},
$$
and under which the fields transform as
\bq
{\bf q}(\st)&\rightarrow&{\bf q}(\ust),\;\;\;{\bf y}(\st)\rightarrow
{\bf y}(\ust),\nonumber\\
\psi(\st)&\rightarrow&\sqrt{u}\psi(\ust),\;\bar{\psi}(\st)\rightarrow
\sqrt{u}\bar{\psi}(\ust).
\eq
The world sheet action given above is not invariant
under this scaling, unless the
constants $g$ and $\alpha^{2}$ are also scaled according to
\be
g\rightarrow u g,\;\;\alpha^{2}\rightarrow u \alpha^{2}.
\ee
Scale invariance is also violated by the two cutoff parameters needed to
regulate the model. We have already introduced one of them, $a$, the
grid spacing in the $\sigma$ direction; in order to regulate the
integral over ${\bf q}$, we will need also to discretize the coordinate
$\tau$, with a grid spacing $a'$. To preserve scale invariance, at least
formally, these two cutoff parameters must also scale like the coordinates:
\be
a\rightarrow u a,\;\;a'\rightarrow u a'.
\ee

All these definitions are somewhat formal, and there is real danger of
violation of scale and hence of Lorentz invariance.
We will see later that, at least within the
framework of the meanfield approximation, physical quantities, which
are by definition cutoff independent, are also scale invariant. Although
this is not a complete
proof, it provides support for the Lorentz invariance of the
whole approach. We will therefore start with the basic assumption that all the
scale invariant quantities formed out of the parameters $\alpha$, $g$, $a$
and $a'$ are finite (cutoff independent), so that  there is
really only one independent cutoff, $a$ or $a'$, 
 with the scale invariant ratio
\be
a/a'=m^{2}.
\ee
fixed at a finite value.  We also note that $m$, which
has the dimensions of mass, is the only dimensionful scale invariant
 parameter in the
problem.  It is therefore a natural candidate
for a Lorentz invariant mass parameter. We will see later that the slope
of the dynamically generated string is proportional to $1/m^{2}$.

From the  remaining parameters $g$ and $\alpha$, we form two new scale
invariant parameters by
\be 
\bar{\alpha}^{2}=\,\frac{a'^{2}}{a}\alpha^{2},\;\;\bar{g}=\,a'\, g.
\ee
For convenience, we have also made them dimensionless, which can always
be achieved by multiplying by appropriate powers of $m$.
As the cutoff parameters $a$ and $a'$ go to zero, in addition to $m$,
 we will keep $\bar{g}$ and
$\bar{\alpha}$ 
fixed at a finite values. If we can then show that physical quantities
can be expressed solely in terms of these three parameters,  their
finiteness and scale invariance will follow.

Before closing this section, we would like to note that in addition to the
coupling constant $\bar{g}$, an extra parameter $\bar{\alpha}$ appeared.
By paying attention to the integration measure in the version of the
model with discretized coordinates, it is possible to argue that this
parameter should be fixed at
$$
\bar{\alpha}^{2}= \frac{1}{4 \pi}.
$$
However, since the exact value of $\bar{\alpha}$ will not of any
importance in what follows, we will leave it undetermined.

\vskip 9pt

\noindent{\bf 3. The Meanfield Expansion}

\vskip 9pt

The idea behind the meanfield approximation is to recognize that Eq.(10)
is the action for a vector model, which can be solved in the large $D$
limit. One first introduces new fields for the scalar products such as
${\bf y}\cdot\dot{{\bf q}}$ and ${\bf y}^{2}$ in order to cast the action
into an expression proportional to $D$, the dimension of the transverse
space, and then the large $D$ limit is taken by the saddle point method [11].
Here, we choose instead an equivalent but simpler approach. Let us add
the following term to $S$ of Eq.(10):
\be
\Delta S=\intst\left(D\kappa\Big(\frac{1}{2}\bar{\psi}(1-\sigma_{3})\psi
-\rho\Big)\right).
\ee
This term merely enforces the definition of $\rho$ through the Lagrange
multiplier $\kappa$. We will use the meanfield method to compute the
ground state expectation value
$$
\rho_{0}=\langle \rho\rangle
$$
of $\rho$. It was argued in references [2-7] that a non-zero value for
$\rho_{0}$ signals a new phase of the underlying field theory, where the solid
lines (boundaries) form a condensate on the world sheet. In contrast,
the original perturbative phase corresponds to $\rho_{0}=0$.

The meanfield calculation is simplified by choosing a configuration of the
world sheet with the $\sigma$ coordinate compactified with periodic
boundary conditions at $\sigma=0$ and $\sigma=p^{+}$,
 and the total transverse momentum ${\bf p}$ set
equal to zero. This configuration has the advantage of being translation
invariant in both $\sigma$ and $\tau$ directions, so that both $\rho_{0}$
and
$$
\kappa_{0}=\langle\kappa\rangle
$$
can be taken to be constants independent of $\st$. Replacing $\rho,\kappa$
by their constant
 expectation values in Eq.(10), the integration over ${\bf y}$ is
easily done, with the result,
\bq
S+\Delta S&\rightarrow&\tilde{S}_{q}+S'+\tilde{S}_{f},\nonumber\\
\tilde{S}_{q}&=& \intst \left(-\frac{1}{2}{\bf q}'^{2}+
\frac{1}{2}\beta^{2} \dot{{\bf q}}^{2}\right),\nonumber\\
\tilde{S}'&=&\intst\left(-D \kappa_{0} \rho_{0}\right),\nonumber\\
\tilde{S}_{f}&=&\intst\left(\bar{\psi}\Big(i\partial_{\tau} -D g \sigma_{1}
+ \frac{1}{2} D \kappa_{0}(1-\sigma_{3})\Big)\psi\right),
\eq
where
\be
\beta^{2}=\frac{\rho_{0}^{2}}{ \alpha^{2} \bar{\rho}_{0}}.
\ee
It is clear that a non-zero value for $\rho_{0}$ results in a
non-zero string slope $\alpha'$:
\be
\alpha'^{2}= \beta^{2}/\pi^{2},
\ee
provided that, as we shall later show,
$$
\bar{\rho}_{0}=\frac{1}{a}- \rho_{0}
$$
is positive, which is necessary for this equation to make sense. This simple
argument shows that, independent of any approximation scheme, a non-zero
value for $\rho_{0}$ leads to string formation. Whether $\rho_{0}$ is 
non-vanishing is a question of dynamics. We will compute below the energy
of the ground state of the system, using the meanfield approximation, and
we will find that it is minimized by a non-vanishing value of $\rho_{0}$.
The energy of the ground state $E_{0}$, which follows from the action given by
Eq.(17), consists of three pieces:
\be
E_{0}=E_{q}+E'+E_{f},
\ee
where $E_{q}$, $E'$ and $E_{f}$ are the contributions of $\tilde{S}_{q}$,
$\tilde{S}'$ and $\tilde{S}_{f}$ respectively. Since $\kappa_{0}$ and
$\rho_{0}$ are constants,
\be
 E'= D p^{+}\kappa_{0} \rho_{0}.
\ee
It is also easy to compute $E_{f}$. After discretizing $\sigma$ in steps of
 length $a$, the system represented by $\tilde{S}_{q}$ consists of
$N=p^{+}/a$ number of decoupled
 one dimensional Ising models. Diagonalizing the resulting two by two
matrices, we have
\be
E_{f}^{\pm}=\frac{D p^{+}}{2 a}\left(\kappa_{0}\pm \sqrt{\kappa_{0}^{2}
+4 g^{2}}\right).
\ee
Clearly, the ground state energy is minimized by taking the minus sign
in front of the square root, so
we will drop the plus sign from now on.

Now consider $E_{q}$.
We shall only need the leading singular (cutoff dependent) part of $E_{q}$,
 in the limit $a, a'\rightarrow 0$,
 with $a/a'=m^{2}$ kept fixed and finite. In the next section, we will show
that, in this limit
\be
E_{q}\rightarrow\frac{D p^{+}}{a'^{2}} F(\beta^{2},m^{2}),
\ee
where $\beta$, (Eq.(18)), is cutoff independent (finite). It is
important to establish the finiteness of $\beta$, since the string slope
$\alpha'$, which we expect to be finite, is proportional to $\beta$. We could
argue this indirectly by observing that $\beta$ is scale invariant; however,
it is worthwhile to carry out an explicit calculation.
 We first note that,
from its definition on a grid with spacing $a$, $\rho_{0}$ has a 
cutoff dependence of the form $1/a$. It is therefore natural to define
\be
\rho_{0}= x/a,\;\;\bar{\rho}_{0}=(1-x)/a.
\ee
 Going back to $S_{f}$ (Eq.(6)), it is easy to show that $x$ represents the
average probability of finding a spin down fermion on the world sheet, which
is also the probability of finding a solid line. Therefore,
$$
0\leq x\leq 1,
$$
Expressed in terms of $x$ and $\bar{\alpha}$ (Eqs.(18) and (24)),
 $\beta$ now reads
\be
\beta=\frac{x}{\bar{\alpha} m^{2}\sqrt{(1-x)}}.
\ee
Since the parameters $\bar{\alpha}$, $x$ and $m$ are all finite
as $a,a'\rightarrow 0$, it follows that $\beta$ is finite, assuming that
$x\neq 1$. From
Eq.(19), the string slope also comes out to be finite. This provides
an after the fact justification for taking $m^{2}$, the ratio of two
cutoffs, to be finite.

Putting together eq.(17),(22) and (23), we have the following expression for
the ground state energy:
\be
E_{0}=D p^{+}\left(\frac{1}{a'^{2}} F(\beta^{2}, m^{2})
-\frac{1}{a} x \kappa_{0}+\frac{1}{2 a}\Big(\kappa_{0}
-\sqrt{\kappa_{0}^{2}+4 g^{2}}\Big)\right),
\ee
and the ground state is determined by the saddle point equations
\be
\frac{\partial E_{0}}{\partial \kappa_{0}}=0,\;\;
\frac{\partial E_{0}}{\partial x}=0.
\ee
The first equation gives
\be
x=\frac{1}{2}\left(1-\frac{\kappa_{0}}{\sqrt{\kappa_{0}^{2}
+4 g^{2}}}\right),
\ee
and eliminating $\kappa_{0}$ in favor of $x$ from this equation,
the ground state energy can be rewritten solely in terms of $x$
($\beta$ is a function of $x$ from Eq.(25)):
\be
E_{0}
=\frac{D p^{+}}{a'^{2}}\left( F\Big(\beta^{2},  
 m^{2}\Big)- \frac{1}{m^{2}} \sqrt{x - x^{2}}
\right).
\ee
To determine $x$ from the second equation in (27), we need to know
the explicit form of $F$. As might be expected, $F$ depends on the
cutoff scheme used.
 We will compute $F$ in the next section, using the cutoff scheme we have
introduced, but we will also consider a large class of related methods
of regulation. The idea is to see to what extent the results derived here
depend on the details of the cutoff scheme. We will show that the ground
state energy given by Eq.(29) above has always a minimum at some value of
$x=x_{m}\neq 0$,  for a large class of cutoff schemes. 
 However, the precise value of $x_{m}$ depends on the scheme
used.
We therefore conclude that $x_{m}\neq 0$ is a
robust result, and string formation with non-zero slope 
takes place. On the other hand, the parameter $\beta$ (Eq.(25))
 and therefore the string slope depends on $x_{m}$. This should not
 really cause concern, since we can argue as follows:
We  take $\alpha'$ or $\beta$ as a given physical parameter, and
for any given value of $x_{m}$, adjust the combination
$$
\bar{\alpha} m^{2}
$$
so as to satisfy Eq.(25). From this point of view, which is the standard
idea behind renormalization, $\bar{\alpha}$, $m^{2}$ and $x_{m}$ are
all  parameters that depend on the cutoff scheme in use; only the string
slope is scheme independent.

\vskip 9pt

\noindent{\bf 4. Cutoff Dependence Of The Ground State Energy}

\vskip 9pt

In this section, we will 
investigate the cutoff dependence of the
ground state energy, first using two sharp cutoffs, and then replacing
 these by two smooth profile functions. In particular, we want to see
whether, independent of the cutoff scheme used, the potential has a
minimum at a non vanishing value of $x$.
 Integrating over ${\bf q}$ in $\tilde{S}_{q}$  (Eq.(17)) gives
\be
\tilde{S}_{q}\rightarrow \frac{1}{2} i D \,Tr\ln\left(\beta^{2}
\partial_{\tau}^{2}- \partial_{\sigma}^{2}\right),
\ee
and the corresponding energy is
\be
E_{q}=\frac{D\,p^{+}}{2 (2\pi)^{2}}\int_{-\Lambda_{0}}^{\Lambda_{0}} d k_{0}
\int_{-\Lambda_{1}}^{\Lambda_{1}} d k_{1} \ln\left(\beta^{2} k_{0}^{2}
+k_{1}^{2}\right).
\ee
Here, we have written the expression for the energy in momentum space, and
we have replaced the grid sizes $a'$ and $a$ by the equivalent momentum
space cutoffs
$$
\Lambda_{0}=\pi/a',\;\;\Lambda_{1}=\pi/a.
$$
Also, strictly speaking, instead of an integral over $k_{1}$, we should have
a discrete sum in steps of $\Delta k_{1}= 2\pi/p^{+}$. However, since we are
only interested in the leading cutoff dependence of $E_{q}$, we can safely
replace this sum by an integral.

As explained earlier, we would like to generalize the sharp cutoff used
above. Accordingly, we replace Eq.(31) by
\be
E_{q}\rightarrow \frac{D\,p^{+}}{2 (2\pi)^{2}}
 \int d k_{0}\int d k_{1}\, f_{0}(k_{0}/\Lambda_{0})
f_{1}(k_{1}/\Lambda_{1}) \ln\left(\beta^{2} k_{0}^{2}+ k_{1}^{2}\right),
\ee
where the profile functions satisfy the following conditions:\\
a) $f_{0}(y)=f_{1}(y)=1$ for $-1\leq y \leq 1$.\\
b) Both $f_{0}(y)$ and $f_{1}(y)$ are 
 positive functions
that rapidly go to zero  as $|y|\rightarrow\infty$. Otherwise, they are
arbitrary.

Let us now extract $F$ from $E_{q}$ according to Eq.(23), by considering the
 limit $\Lambda_{0,1}\rightarrow \infty$, with $\Lambda_{0}/\Lambda_{1}
=m^{2}$ fixed. The result is
\be
F=\frac{1}{8 m^{2}}\int d y \int d z f_{0}(y) f_{1}(z)\left(
\ln(m^{2} \beta^{2} y^{2}+ z^{2})+\ln(\Lambda_{1})\right).
\ee
The term proportional to $\ln(\Lambda_{1})$ on the right hand side is
independent of $x$ ($\beta$) and therefore
 it does not contribute to the equations (27)
that determine the ground state. In fact, it is an irrelevant constant which
can be subtracted from the ground state energy and we will drop it from now
on.

To prove that $E_{0}$ has a minimum at $x_{m}\neq 0$, we have to first
show that $F$ is an increasing function of $x$. From the above equation, 
it is clear that
$$
\partial F/\partial \beta >0,
$$
and from Eq.(25),
$$
\partial\beta/\partial x>0,
$$
and therefore
\be
\partial F/\partial x>0.
\ee
For small enough $x$, we can approximate $F$ by
\be
F\approx C_{0}+ C_{1} x,
\ee
where the constant 
$$
C_{1}=(\partial F/\partial x)_{x=0}
$$
is positive. On the other hand, the second term in Eq.(29) for $E_{0}$ is
negative, and for small $x$, it can be approximated by
$$
-\sqrt{x -x^{2}}/m^{2}\approx -\sqrt{x}/m^{2}.
$$
Comparing this with (35), we see that for small enough $x$,
the second term dominates, and $E_{0}$ is a decreasing function of $x$.
On the other hand, for $x>1/2$, the second term now becomes an increasing
function, and consequently, so does $E_{0}$. Since $E_{0}$ is decreasing
for small $x$ and starts increasing at $x=1/2$, it must have a minimum
for some $x_{m}$ in the interval $0<x \leq 1/2$.

We note that the only property of $f_{0,1}$ needed in the above argument is
their positivity. Therefore, we believe that the result is quite
 general, and that it holds in any 
regulator scheme which does not violate the positivity properties of the
underlying field theory. Hence, string formation is a robust result,
which does not depend on the details of the regulation scheme used.
 Although the existence of $x_{m}\neq 0$ is thus
assured, its value clearly depends on the cutoff functions $f_{0,1}$. We
have argued at the end of the last section that $x_{m}$, $\bar{\alpha}$ and
$m^{2}$ are all scheme dependent parameters; only the physical parameter
built out of them, $\alpha'$, need be scheme independent.

Before closing this section, we would like to mention that
 the regulation scheme used
in reference [3-7] corresponds to taking the small $x$ and therefore
the small $\beta$ limit in Eq.(33), which gives the simple result,
$$
F\approx \frac{x}{4 \pi \bar{\alpha} m^{2}} \Lambda_{0}^{2}.
$$

\vskip 9pt

\noindent{\bf 5. Sigma Model For $\phi$  }

\vskip 9pt

In this section, we will study higher order corrections to the leading mean
 field approximation. This will involve evaluating $\tilde{S}_{q}$ (Eq.(30)),
not just for a constant $\beta$, as was done in the last section, but with
$\beta$ taken to be an arbitrary function of the coordinates. Of course, this
cannot be done exactly, so we resort to a double power series expansion
described below. Actually, we will explicitly evaluate only the first two terms
of this expansion. As for the rest of the terms, we will only be interested
in their dependence on the cutoff. We first rewrite Eq.(30) as
 \be
\tilde{S}_{q}
 \rightarrow \frac{1}{2} D\,Tr\ln\left(-\partial_{\tau}
\phi \partial_{\tau}- \partial_{\sigma}^{2}\right),
\ee
where we have renamed  $\beta^{2}$ as  $\phi(\st)$, and switched
to Euclidean metric on the world sheet. From now on, the world sheet
action will be written using the Euclidean metric. Next, we expand
$\tilde{S}_{q}$ in powers of derivatives with respect to the world 
sheet coordinates acting on $\phi$: 
\be
\tilde{S}_{q}= \intst \left(
 U_{q}(\phi)+ \sum_{s=1}^{\infty}
Z_{q}^{(s)}(\phi)\, \partial^{2s}\phi \right).
\ee
To simplify the notation, we have written the above expansion in a
somewhat schematic
form. The derivatives are with respect to $\sigma$ or $\tau$, and
 they could be acting on several $\phi$'s instead a single one as shown.  
Therefore, in reality, several terms  are collectively represented by a single
$Z_{q}^{(s)}$. The  common feature of these terms is that they go
with $2 s$ number of derivatives, which is the only feature relevant
in the following power counting arguments.

This  well known expansion is frequently used in the effective potential
computations. In conjunction with the loop expansion, it makes the calculation
of at least the leading terms in the expansion possible [10]. From
our point of view, its chief advantage is that it keeps track of the
cutoff dependence of the action. By  naive power counting, one would
expect each derivative  to go with an inverse power of the
cutoff. This is true, of course, up to factors involving powers of the
logarithm of the cutoff. These logarithmic factors,
which will be the focus of our attention, will play an important role
later on.

 In order to make progress, we need a second expansion, this time in
powers of $1/D$. This is the same $1/D$ expansion on which the mean
field method is based, and it is accomplished by
first splitting $\phi$ into two parts as
$$
\phi=\phi_{0}+\chi,
$$
and then expanding the trace in Eq.(36) in powers of $\chi$:
\be
Tr\ln\left(-\partial_{\tau}
\phi \partial_{\tau}- \partial_{\sigma}^{2}\right)=
\sum_{0}^{\infty}\frac{(-1)^{n}}{n+1} Tr\left(-W(\Delta^{-1} W)^{n}\right),
\ee
where,
$$
\Delta=\phi_{0}\,\partial_{\tau}^{2}+\partial_{\sigma}^{2},\;\;
W=\partial_{\tau} \chi \partial_{\tau}.
$$
For convenience of calculation, $\phi_{0}$ is taken to be a
 constant ($\st$ independent) background field. We could set it
equal to the ground state expectation value of $\phi$; however,
we will keep it arbitrary for the time being. The general term in the
expansion is the one loop graph with $n+1$ external lines shown in
Fig.3.
\begin{figure}[t]
\centerline{\epsfig{file=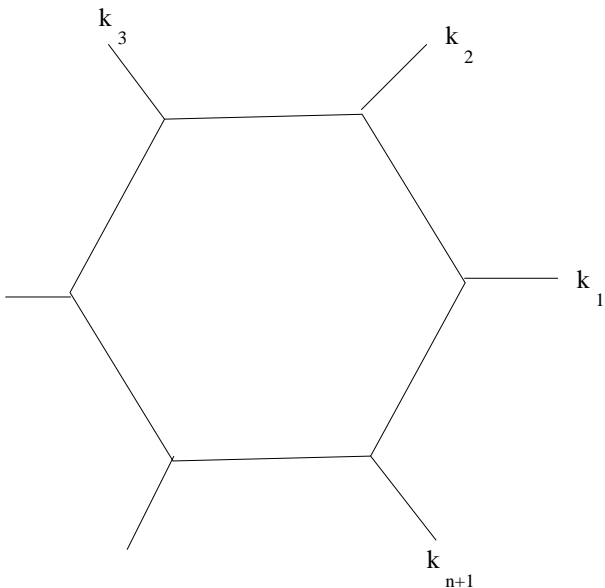,width=8cm}}
\caption{One Loop Graph}
\end{figure}

The next step is to translate the derivative expansion into the 
momentum space, by expanding the graph in Fig.3 in powers of
$k_{1}, k_{2},\cdots, k_{n+1}$ around the point $k_{1}=k_{2}=\cdots=
k_{n+1}=0$. Here, the $k$'s are two dimensional world sheet momenta
carried by the external lines in Fig.3. The term proportional to
$Z^{(s)}_{q}$ in Eq.(37) corresponds to a term with
the total power $2 s$ in $k$'s in this expansion. 
Let us start with the first term of this
expansion where all the momenta are set equal to zero; this corresponds
to $U_{q}$ in Eq.(37). Setting all the momenta equal to zero is the same as
replacing the $n$'th power of $\chi$ in the expansion (38) by
$$
\intst\left(\chi(\st)^{n}\right).
$$
This leads to the following recipe: Compute
$$
\frac{1}{2} Tr\ln\left(-\phi_{0}\, \partial_{\tau}^{2}-\partial_{\sigma}^{2}
\right)
$$
 for a
constant $\phi_{0}$, and in the result
 replace $\phi_{0}$ by $\phi(\st)$ and then integrate
over $\st$. The calculation with a constant background was already done in the
last section, with $\beta^{2}$ replacing $\phi_{0}$. The result can be 
expressed in terms of $F$ (Eq.(33)):
\be
U_{q}(\phi(\st))= \frac{D  \Lambda_{0}^{2}}{\pi^{2}}
F\left(\phi(\st), m^{2}\right).
\ee
Here we note the quadratic dependence on the cutoff, which is all we need
in the future. Also, the distinction between $\Lambda_{0}$ and
$\Lambda_{1}$ will not play any role in our subsequent discussion.
 We will  simplify writing by replacing 
$\Lambda_{0,1}$ by a generic cutoff $\Lambda$ from now on.

Next we turn our attention to $Z_{q}^{(1)}$ , which is the term
corresponding to $n=1$ in the series in Eq.(37). (The term $n=0$ vanishes).
The corresponding graph is Fig.3 with two external lines.
 After evaluating this graph in
the presence of a constant background $\phi_{0}$, we expand to second
order in the external momentum $k$, and then set $k=0$. This provides the
two derivatives acting on the two external $\chi$ fields. We can then replace
 the two $\chi$'s by $\phi$'s, since these two differ only by a constant.
Furthermore, repeating the  argument we have used in the case of $V_{q}$,
after the graph has been evaluated, we can replace $\phi_{0}$
by $\phi(\st)$. The result, which was
computed in references [5-7], has a logarithmic dependence on the cutoff:
\be
\intst\left(Z_{q}^{1} \partial^{2}\phi\right)
=\frac{D}{64 \pi} \ln(\Lambda/\mu)\intst
\,\phi^{-5/2} \left(
\phi\, (\partial_{\tau}\phi)^{2}+(\partial_{\sigma}\phi)^{2}\right).
\ee
A few comments are in order:\\
a) This expression involves  the factor $\phi^{-5/2}$, which only makes sense
if we agree to expand it around a constant background $\phi_{0}$. In the
last section, we have identified $\phi_{0}$ with the ground state
expectation value of $\phi$, and determined it by minimizing
$U(\phi_{0})$. Although, as we have seen in section 4,
 the particular value of $\phi_{0}$ is somewhat dependent on the
details how the model is regulated, the important point is that it does not
vanish. Since, from Eqs.(18) and (19),
$$
\alpha'^{2}\rightarrow \phi_{0}/\pi^{2},
$$
 a non-zero $\phi_{0}$ means a non-zero string
slope.  We also note that a non-zero $\phi_{0}$, in addition to a non-zero
 slope, generates a
kinetic energy term for $\phi$, and $\phi$, which was originally an
auxilliary field,  becomes dynamical. In what follows, so long as it is
non-zero, the particular value of $\phi_{0}$ is really not important.\\ 
b) $\mu$ is an infrared cutoff, needed to avoid  infrared
divergence. Since, so far, we were only interested in the dependence on the
ultraviolet cutoff, the infrared cutoff did not matter.
 Later, when we reexamine the problem from the point of view
of the renormalization group, the basic idea will be to lower
the ultraviolet cutoff $\Lambda$ to a lower value, and $\mu$
will be identified with this lower cutoff.\\
c) By dimensional reasoning, it is easy to show that
 terms corresponding to $s>1$ in the expansion  (37) go with inverse
powers of cutoff:
\be
Z_{q}^{(s)}\sim \Lambda^{2- 2 s}.
\ee
Therefore, one may suspect that these terms can be ignored.
Actually, this is misleading; the power suppression is compensated
by the higher number of derivatives. Later on, we will see that
that these terms are suppressed not by powers but by factors
logarithmic in $\Lambda$.\\
d) We  note that $s=1$ is a special case. That is the only term in the
expansion that has a logarithmic dependence on the cutoff (see Eq.(40)):
\be
Z^{(1)}_{q}\sim \ln(\Lambda/\mu).
\ee

So far, we have been studying $\tilde{S}_{q}$, the contribution of the field
${\bf q}$ to the action. To this must be added
the contribution of the fermionic sector, represented by $\tilde{S}'$ and 
$\tilde{S}_{f}$ in Eq.(17) to arrive at the total action. After integrating
over the auxilliary fields $\kappa$, $\psi$ and $\bar{\psi}$, the
resulting effective action, $S_{e}$,
 which now depends only on $\phi$, can be expanded
in powers of spatial derivatives just as in Eq.(37):
\be
S_{e}=\intst\left(U(\phi)+\sum_{s=1}^{\infty}
Z^{(s)}(\phi) \partial^{2 s}\phi \right).
\ee
We have already computed $U(\phi)$; it is essentially given by the right hand
side of Eq.(29):
$$
U= \frac{\Lambda^{2}}{\pi^{2}}\left(F(\phi,m^{2})- \frac{1}{m^{2}}
\sqrt{x- x^{2}}\right),
$$
where $x$ is the solution to (see Eq.(25))
$$
\phi=\frac{x^{2}}{\bar{\alpha}^{2} m^{4}(1-x)}.
$$
 Note  that
just as $U_{q}$, $U$ is quadratic in the cutoff $\Lambda$. The cutoff
 dependences of $Z^{(s)}$ for $s>1$ are also unchanged; they are given
by simple power counting, as in Eq.(41). Also, a simple
calculation shows that there is no contribution to $Z^{(1)}$ from the
fermionic sector.
To sum up:\\
a) The terms in Eq.(43) have the same cutoff dependence as in
Eqs.(41,42): $U$ is proportional to $\Lambda^{2}$,
 and $Z^{(s)}$ is proportional
to $\Lambda^{2- 2 s}$ for $s>1$.\\
b) $Z^{(1)}=Z_{q}^{(1)}$ is again given by Eq.(40); and so its cutoff dependence
 is logarithmic. 

\vskip 9pt

\noindent{\bf 6. Transforming The Action   }

\vskip 9pt

In this section, we will try to address the following question: How does one
work with an action of the form given by Eq.(43), which is both explicitly
cutoff dependent, and also formally non-renormalizable, since it
contains terms with more than two derivatives?
Of course, here we have a somewhat unusual situation;
the effective action $S_{e}$ was
gotten by doing a one loop calculation, so one might think that the
cutoff dependence could be eliminated by introducing counter terms.
These counter terms usually arise from adding cutoff dependent
pieces to the adjustable parameters (masses, coupling constants) in
the theory. The adjustable parameters in our
problem are $\alpha$ and $g$, and their dependence on the cutoff was 
already fixed by Eq.(15). We recall that Eq.(15) was forced on us
by requiring the invariance of the original action  under
the scaling of the world sheet coordinates. We have also stressed
that  scaling invariance follows from invariance under a special
boost, and it cannot be violated without breaking Lorentz invariance. Since
there is no freedom of tuning the cutoff dependence of $\alpha$ and $g$,
one might consider introducing  $\phi$ dependent counter terms
 into $S_{e}$ to cancel, for example,
 the cutoff dependent part of $U$. One must
remember, however, that $\phi$ (or $\beta^{2}$) was a non-dynamical
auxilliary field in the original action of Eq.(10), and the dependence
of the action on $\phi$ came from $\Delta S$ and it was completely fixed. 
It has later acquired a kinetic
energy term (Eq.(40)) from quantum effects and  as a result became dynamical.
Therefore, 
given the initial action, the  $\phi$ dependence of
 $S_{e}$ is completely fixed. Towards the end of this section, we will
show that by suitably scaling both the world sheet coordinates and the
field $\phi$, the action can be transformed into a form that is a 
suitable starting point for a perturbation expansion. 

A second problem  is how
to identify the set of observable Greens functions required to be cutoff 
independent. We will make the reasonable assumption that physics
lies in the Lorentz invariant sector of the model, and therefore we
should only demand that Lorentz invariant Greens functions should
be free of the cutoff. We recall that,
 in contrast to the covariant approach to field 
theory, we do not here have manifest Lorentz invariance. In addition,
the cutoffs $a$ and $a'$ (or $\Lambda_{0,1}$) 
 are not Lorentz scalars. In view of this, it is too much to demand that
all Greens functions be finite: The cutoff dependendence may be an artifact
due to non-covariance, and it may disappear from the Lorentz invariant 
sector of the model. 

This brings up the question of how to identify Lorentz invariant objects.
Investigation of full Lorentz invariance is a difficult task which will not
be attempted in this paper. Instead, we will
construct  Greens functions invariant
under a subgoup of the Lorentz group consisting of rotations in the
transverse directions and boost along the special light cone direction.
As we have stressed earlier, the special boost is realized as scaling of the
world sheet coordinates, and so we should be looking for scale invariant Greens
functions. These are easy to construct, for example, the composite fields
\be
\intst\, G(\phi)\,(\partial\phi)^{2},\;\;\;\intst\, E(\phi)\partial{\bf q}
\cdot \partial{\bf q},
\ee
where $G$ and $E$ are arbitrary functions and $\partial$ is derivative
with respect to $\sigma$ or $\tau$, are scale invariant.  
 The Greens functions constructed from the products
of these composite fields will also share this property. It is easy
to construct many more examples.  Note that these examples are all
non-local: Since the
world sheet coordinates transform under scaling, one has to integrate
over them to form scale invariant quantities.

We will now argue that scale invariant objects are also cutoff
independent, at least if we focus on power dependence and ignore
logarithmic corrections. This follows because the only parameter in the
problem that scales non-trivially is the cutoff $\Lambda$; under the scaling
$$
\sigma\rightarrow u \sigma,\;\;\tau\rightarrow u \tau,
$$
$\Lambda$ transforms as 
$$
\Lambda\rightarrow \Lambda/u.
$$
Therefore, the  scaling dimension of any dyanamical variable must solely
come from its dependence on the cutoff. For example, in the derivative
expansion of Eq.(43), the cutoff dependences of $U$ and $Z^{(s)}$, given by
Eq.(41), all follow from their scaling dimensions. As a special case,
scale invariance implies cutoff independence.

The above discussion is incomplete in two important respects: First,
it was completely classical; for example, we ignored the logarithmic
dependence of $Z^{(1)}$, which is purely quantum in origin.
Also, we have so far only considered the  contributions coming
from integrating over ${\bf q}$; we should also take into account
the contributions coming from
integrating over $\phi$, since the latter has become a
 full fledged dynamical field after the ${\bf q}$ integration.
In the next section, we will take care of these omissions by 
considering a perturbation expansion, starting with $S_{e}$.
The expansion parameter will initially be 
   $1/D$, the natural small parameter in the large $D$ limit, although
it will be replaced by a more appropriate one later on. In 
preparation for the perturbation expansion, it is convenient to
transform the action so that the free part (mass and kinetic energy)
does not depend on the cutoff or on $D$. We first write $\phi$ as the
sum of its expactation value $\phi_{0}$ and a fluctuating field which
we called $\chi$ in Eq.(38). In order not to complicate the notation, instead
of introducing a new field  $\chi$, we will simply make the replacement
\be
\phi\rightarrow \phi_{0}+ \phi.
\ee
After this replacement, we will rescale $\phi$ by
\be
\bar{\phi}=\left(D\ln(\Lambda/\mu)\right)^{1/2}\,\phi,
\ee
and combine it with a change of the world sheet coordinates
 from $\st$ to $\bar{\sigma},\bar{\tau}$:
\be
\bar{\sigma}= \Lambda \left(\ln(\Lambda/\mu)\right)^{-1/2}\,\sigma,\;\;\;
\bar{\tau}=\Lambda \left(\ln(\Lambda/\mu)\right)^{-1/2}\,\tau.
\ee

After these transformations, $S_{e}$ can be written as
\be
S_{e}=\int d\bar{\sigma}
 \int d \bar{\tau}\left(\bar{U}(\bar{\phi})
+\sum_{1}^{\infty}\bar{Z}^{(s)}(\bar{\phi})\,\partial^{2 s}\bar{\phi}
\right),
\ee
where the barred and unbarred quantities are related by
\be
\bar{U}=\Lambda^{-2}\,\ln(\Lambda/\mu)\,U,\;\;
 \bar{Z}^{(s)}=\Lambda^{2s -2}\,
\left(\ln(\Lambda/\mu)\right)^{1-s} \,Z^{(s)}.
\ee

To see more clearly what this transformation has accomplished,
it is useful to expand the barred quantities in powers $\bar{\phi}$ by
\be
\bar{U}=\sum_{2}^{\infty}\bar{U}_{n}\,\bar{\phi}^{n},\;\;
\bar{Z}^{(s)}=\sum_{1}^{\infty}\bar{Z}^{(s)}_{n}\,\bar{\phi}^{n},
\ee
and  check the dependence
of these terms on the cutoff and on $D$. Defining
\be
t=\ln(\Lambda/\mu),
\ee
 and combining the cutoff dependence given by Eqs.(41) and (42) with the
transformations described above,
 we see that
\be
\bar{U}_{n}\sim (D\,t)^{1- n/2},
\ee
and,
\be
\bar{Z}^{(1)}_{n}\sim (D\,t)^{(1-n)/2},
\ee
for $s=1$ and
\be
\bar{Z}^{(s)}_{n}\sim (D\,t)^
{(1 -n)/2}\,t^{-s},
\ee
for $s>1$. The reason for a different dependence on  $t$
for $s=1$ can be traced back to 
$Z^{(1)}$, which is the only term in the expansion proportional to
$t$.

From the above results, we see that:\\
a) There is no longer any dependence on powers of $\Lambda$
 in the new
action. The remaining cutoff dependence is purely logarithmic.
One can directly see from Eqs.(51-54) that the old cutoff $\Lambda$
has been replaced by the new cutoff $t$.\\
b) The free part of the action, given by
\be
S_{e}^{(0)}=\int_{0}^{p^{+}} d\bar{\sigma}\int d \bar{\tau}
\left(\bar{U}_{2}\, \bar{\phi}^{2}+ \bar{Z}_{1}^{1}\, \bar{\phi}\,
 \partial^{2}\bar{\phi}\right),
\ee
is both cutoff and $D$ independent. We recall that this was the main
 motivation for transforming the action. Identifying $D\,t$ with the
wave function renormalization constant, we note that the scaling
of $\phi$ in Eq.(46) very much looks like the standard
 wave function renormalization of field theory. There is, however, an
important  difference: Wave function renormalization constant normally
looks like
$$
Z\sim 1+ e^{2}\, \ln(\Lambda/\mu),
$$
whereas, with the identification $e^{2}=1/D$, we have
$$
Z\sim e^{-2}\,\ln(\Lambda/\mu).
$$
This is because, in our case, the kinetic energy term for $\phi$ is
not present in the original action, but it is generated at the
level of one loop.\\
c) Examining Eqs.(52,53), we see that, instead of $1/\sqrt{D}$,
 it is more natural to identify
\be
\lambda=1/\sqrt{D\,t}
\ee
as the parameter of the perturbation expansion. This can be 
made more precise by defining $\lambda$
by setting the coefficient of the term with $s=1$, $n=2$ equal to
$\lambda$, so that, up to an irrelevant multiplicative
constant, it appears in the action in the form
\be
\int d\sigma \int d\tau \left(\lambda\,\bar{\phi}^{2}
\partial^{2}\bar{\phi}\right).
\ee

With this new identification of the coupling constant, the convergence
of the expansion around the meanfield is no longer tied to the number
of transverse dimensions being large: $\lambda$ could be small without
$D$ being large if $\Lambda/\mu$ is large. Later on, we will see that
one can identify $\lambda$ as the running coupling constant defined
at scale $\mu$. As $\mu\rightarrow 0$, the model tends towards a free
field theory. There are, however, obstacles in the infrared
 that prevent us from reaching $\mu=0$.
For one thing, we have decompactified the sigma coordinate and replaced
a discrete spectrum by a continuous one. This approximation, valid in the
ultraviolet region, breaks down when we approach the infrared. In addition,
there is a problem with a strongly coupled zero mode in the spectrum [11].
Also, by just dimensional reasoning, the $\mu$ dependence of the $Z$'s
is given by
$$
Z^{(s)}\sim \mu^{2 s-2}
$$
for $s>1$, so there is a serious infrared divergence. All we can say is that,
as we reduce $\mu$, at least for a while, $\lambda$ tends to
decrease. We will make this point a bit more precise when we discuss the
renormalization group.

Now we are ready to give at least a partial answer to the question
posed at the beginning of this section. We have constructed a new action, 
Eq.(48), in which the original cutoff $\Lambda$ is replaced by
the logarithmic cutoff $t$.
 For computing Lorentz and
hence scale invariant quantities, it can be used in place of the
old action of
Eq.(43), since the two are related by a scale transformation. The free part
of the action is cutoff independent, and the only cutoff dependence 
appears in the interaction. If formally the limit $t\rightarrow \infty$
is taken, the theory becomes free.  Of course, one still have to
deal with the additional cutoff dependence coming from higher order terms,
 but at least, there is now a satisfactory starting point, eq.(48),
 for doing the perturbation expansion.

\vskip 9pt

\noindent{\bf 7. Perturbation Expansion And The Renormalization Group    }

\vskip 9pt

 Our first goal
is to determine the dependence of the individual terms
of the perturbation expansion on the cutoff
 $t$ for large $t$. Later we will use the knowledge thus gained 
 to introduce an approach to the problem based
on the renormalization group.
 We have already noted that, for large $t$,
the expansion parameter $\lambda$ is small, and so the perturbative
treatment is justified. 
The tool we will use to investigate the large $t$ dependence
is elementary power counting.
 Let us first introduce a uniform
notation for interaction vertices by defining
$$
W^{(0)}_{n}=\bar{U}_{n},\;\;W^{(s)}_{n}=\bar{Z}^{(s)}_{n+1},\;\;
s\geq 1,\;n>2.
$$
The subscript $n$ on $W$ counts the number of lines meeting at the
vertex, and $2 s$ counts the total number of derivatives (momenta)
associated with various lines. The asymptotic
$t$ dependence of the $W$'s can
be read from Eqs.(52-54):
\be
W_{n}^{(s)}\sim t^{(1-s-n/2+\delta_{1,s})}.
\ee

Now consider a contribution to
the vertex $W^{(S)}_{E}$ from a graph with $E$ external lines,
$I$ internal lines, $V$ vertices
 and $L$ loops (unconstrained momenta).
 Very
schematically, this graph can be represented as
\be
W_{E}^{(S)}\sim \int^{\sqrt{t}}\prod_{i=1}^{L} d^{2} k_{i}
 \prod_{j=1}^{I} (k_{j}^{2}+M^{2})^{-1} \prod_{l,m=1}^{V}
\left((k_{l})^{2 s_{l}- 2 s_{m}}\,W_{n_{l}}^{(s_{l})}
\right),
\ee
where, $n_{l}$ represents the number of lines entering the $l$'th
vertex and $2 s_{l}$ the number of momenta associated with that vertex.
Of these momenta, $2 s_{m}$ of them are associated with external
lines and the rest with internal lines.
 These quantities satisfy the following
constraints:
\be
\sum_{l=1}^{V} n_{l}= E+ 2 I,\;\;\sum_{m=1}^{V} s_{m}=S,\;\;
V=I-L+1.
\ee
In writing Eq.(59), we have suppressed many details; for example,
we have not expressed $k_{l}$'s, the momenta entering the vertices,
 as linear combinations of the internal momenta
$k_{i}$. Also, Lorentz indices have been suppressed. Since we are only
interested in counting powers of $t$, these details do not matter.

To find the large $t$ behaviour of $W_{E}^{(S)}$, we use  Eq.(58) for the
$W_{n}^{(s)}$'s on the right side, and also, we change variables
by
$$
k_{i}\rightarrow \sqrt{t}\,k_{i}.
$$
The result is
\be
W_{E}^{(S)}\sim t^{L-I}\,\prod_{l,m=1}^{V} t^{1-s_{m} -n_{l}/2
+\delta_{s_{l},1}}= t^{L -2 I -S - E/2}\,\prod_{l=1}^{V}
t^{1+\delta_{s_{l},1}},
\ee
where (60) was used in the second step. Clearly,
 the dominant contribution comes from $s_{l}=1$, corresponding to vertices
$W_{n}^{(1)}$. Letting $s_{l}=1$ in the above expression,
and using $V=I-L+1$, we have,
\be
W_{E}^{(S)}\sim t^{L-2 I -S +2 V -E/2}= t^{-L -S -E/2 +2}.
\ee
We see that one loop contribution, $L=1$ dominates, and finally,
the leading $t$ dependence is
\be
W_{E}^{(S)}\sim t^{1- S -E/2}.
\ee
But this is just this the original $t$ dependence given by (58) for
vertices with $S>1$. Therefore,  higher order 
contributions do not change the power dependence of vertices with
$S>1$, although they do contribute to the constant multiplying this
power. In the case of vertices with $S=1$,
the power  given by (63) is down by one unit compared
to the original one  (Eq.(58)), so in this case there is
no correction  to the original dominant asymptotic behaviour.
To summarize:\\
a) For large $t$, the theory is weakly coupled, with the coupling
constant $\lambda$ given by Eq.(56).\\
b) The $t$ dependences of the vertices given by Eq.(58) are unchanged
when higher order perturbation results are taken into account.\\
c) Perturbatively,  the vertices $W_{n}^{(1)}$, with $s=1$, dominate
 the  large $t$ behaviour; $W$'s with $s>1$ are negligible
in comparison. We note that these are the so called non-renormalizable
interaction terms with more than two derivatives. In fact, they are
damped by inverse powers of $t$ and are therefore harmless.\\
d) The $W$'s with $s=1$ themselves receive no higher order corrections. 
These vertices are therefore still given by Eq.(40): One has to expand
 the right hand side of this equation in powers of $\phi$ about a
 non-zero field expectation value.

We will now reexamine the results obtained so far from the
point of view of the renormalization group. The initial model
 was regulated by a grid on the world sheet. The grid spacing, or
equivalently, a corresponding momentum cutoff $\Lambda$ acted as the
regulator. The basic idea of the renormalization group is to lower
the cutoff systematically by integrating over slices of momenta,
finally arriving at an effective action sensitive only to low values
of momenta (energy). Here, we can implement this idea
as follows: We first start with a very large cutoff $\Lambda^{(0)}$,
and integrate over the modes between $\Lambda^{(0)}$ and $\Lambda$
of the field ${\bf q}$, where  the ratio
$\Lambda^{(0)}/\Lambda$
 is taken to be large. The result of this
integration can be schemetically represented by
\be
S(\Lambda_{0})=S_{q}(\Lambda)+S_{e}(\Lambda^{(0)},\Lambda).
\ee
In this equation, $S(\Lambda^{(0)})$ stands for the original action,
with the modes of the field ${\bf q}$ cutoff at $\Lambda^{(0)}$.
After integrating over the modes between $\Lambda^{(0)}$ and $\Lambda$,
and  carrying out the large $D$ expansion, as we have done in sections
3 to 5, we end up with the two terms on the right hand side. The first
term,
\be
S_{q}(\Lambda)=\int d\sigma\int d\tau\left(\frac{1}{2} {\bf q}'^{2}
+\frac{1}{2}(\phi_{0}+
\lambda\bar{\phi)}\,\dot{{\bf q}}^{2} \right)E(\Lambda),
\ee
is the original action for ${\bf q}$, with, however, the cutoff
reduced from $\Lambda^{(0)}$ to $\Lambda$. Here $E(\Lambda)$ is a projection
operator which cuts off the modes of both ${\bf q}$ and $\phi$ at
$\Lambda$. Compared to Eq.(36), we have $\phi_{0}+\lambda\bar{\phi}$
 in place of simply
$\phi$ because of the shift $\phi\rightarrow \phi_{0}+\phi$ (Eq.(45)),
followed by the scaling $\phi=\lambda \bar{\phi}$ (Eq.(46)).

The second term on the right side of the equation is the action for
$\phi$, generated by the integration over the modes of ${\bf q}$
between $\Lambda^{(0)}$ and $\Lambda$. This term can again be expanded as
in Eq.(43). The only difference is in the cutoffs used in computing it:
 The ultraviolet
cutoff is now $\Lambda^{(0)}$ and the infrared cutoff is $\Lambda$. If 
 the transformation to the barred variables  is carried
out, everything works out the same as before, with the exception that,
$t$  is now replaced by 
$$
t\rightarrow \ln(\Lambda^{(0)}/\Lambda).
$$

The reason for starting  with an initial cutoff $\Lambda^{(0)}$
instead of immediately with $\Lambda$
is the following: The field $\phi$ did not have a kinetic energy term
in the initial action, and such a term is needed in order to be able
to integrate perturbatively over this field. Integrating only over the
modes of ${\bf q}$ from $\Lambda^{(0)}$ to $\Lambda$ initially
generates the needed kinetic energy term. The ratio $\Lambda^{(0)}/\Lambda$
is taken to be large so that the coupling constant is small and 
therefore perturbation expansion is valid. Once the kinetic energy
term for $\phi$ is generated, one can take 
$S(\Lambda^{(0)})$ as the starting point of the renormalization
group analysis. Having reduced the cutoff from $\Lambda^{(0)}$ to
$\Lambda$, $\Lambda$ becomes the new ultraviolet cutoff. We can continue
reducing $\Lambda$ further by integrating over the modes of both
${\bf q}$ and $\phi$ together
 till we reach a much lower scale $\mu$.
 Following the conventional treatment, this can be done
in infinitesimal steps, and a set of coupled differential equations
for the $W$'s can be derived. However, here the situation is much simpler;
 we can dispense with this elaborate machinery, relying
instead on the results
from perturbation theory, summarized following Eq.(63).
 As we keep on reducing $\Lambda$ by integrating over the
modes of ${\bf q}$ and $\bar{\phi}$, the coupling constant $\lambda$ is
still given by Eq.(56), with the only difference that the definition
of $t$ keeps changing. If $\Lambda$ is reduced to a new value $\mu$,
the corresponding $t$ is given by
$$
t\rightarrow \ln(\Lambda^{(0)}/\mu).
$$
This is because to the leading order in $t$,
$\lambda$ receives no contribution
from the integration over the modes $\bar{\phi}$ (see item d) of the
summary following Eq.(63)). The only contribution comes from the
integration over the modes of ${\bf q}$, and that simply changes
the lower cutoff in Eq.(51) from $\Lambda$ to $\mu$. As a consequence,
the coupling constant runs towards zero as the lower cutoff tends to zero,
and the model appears to flow towards a free theory in the infrared.
 Of course, we pointed out at the end of section 6, as $\mu$
gets small, at some point the approximations we are making are no
longer valid. It is very likely that at some 
value of $\mu$, the coupling
constant $\lambda$ reaches a minimum, and as $\mu$ is decreased further,
it starts growing. The theory then becomes strongly coupled,
and the mean field method, based on the expansion in $\lambda$,
breaks down.
 
All this is telling us that the mean field approximation is good
when applied to highly excited states of the string, but it
breaks down when applied to the lower lying spectrum.
One can therefore trust it for computing the asymptotic slope, but not
for computing, for example, the intercept. To see this more clearly,
  instead of letting $\mu\rightarrow 0$,
we can let $\Lambda\rightarrow\infty$, while still keeping $\mu$
large. In this limit, $\lambda$ again tends to zero, the fluctuations
of the string slope tend to die out, and to a good approximation,
  we get a free string with a fixed
slope. We note that in the light cone variables we have chosen,
$\mu p^{+}$ corresponds to the square of the total momentum. Therefore,
at large $\mu$, we are probing the higher excitations of the string,
with the square of the mass of the order of $\mu p^{+}$ or larger.
It is only in this regime that we can trust the perturbation
expansion, and the resulting picture of formation of a free string.

\vskip 9pt

\noindent{\bf 8. Conclusions And Open Problems}

\vskip 9pt

In the present paper, we continued  earlier work on the world sheet
approach to field theory and the mean field calculations based on it.
In particular, we focused on the  cutoffs needed for the world sheet set up,
and we investigated  possible cutoff dependences of various results
based on the mean field method. We showed that after integrating over the
target space field ${\bf q}$, a new field $\phi$ was dynamically generated,
with a generalized sigma model for its action.
This action was explicitly cutoff dependent; however, after a simple
transformation, all the cutoff dependence could be absorbed into the
definition of a running coupling constant. An argument based on the
idea of the renormalization group was used to show that, as the cutoff was
lowered, the coupling constant approached zero, and
  the model tended towards a free string. We argued, however, that because
of various infrared problems, it was not possible to really reach this limit.
Nevertheless, by keeping the cutoff sufficiently large, we could still
obtain useful information about the ultraviolet behaviour of the model:
For example, one could conclude that
the higher exited modes were indeed described by a free string. We
consider this as the main result of the present work.

Many  problems, some of them listed at the beginning of the
introduction, such as full Lorentz invariance, still remain open.
Another problem, brought to light here, is the likely breakdown
of the mean field expansion for the low lying states of the string.
It would be very desirable to find a way of extending the mean
field method, so that, for example, the string intercept could be
reliably determined.

\vskip 9pt

\noindent{\bf Acknowledgement}

\vskip 9pt

This work was supported in part
 by the Director, Office of Science,
 Office of High Energy and Nuclear Physics, 
 of the U.S. Department of Energy under Contract 
DE-AC02-05CH11231.

\vskip 9pt

\noindent{\bf References}

\vskip 9pt

\begin{enumerate}
\item K.Bardakci, C.B.Thorn, Nucl.Phys. {\bf B 626} (2002) 287,
hep-th/0110301.
\item K.Bardakci, C.B.Thorn, Nucl.Phys. {\bf B 652} (2003) 196,
hep-th/0206205.
\item K.Bardakci, C.B.Thorn, Nucl.Phys. {\bf B 661} (2003) 235,
hep-th/0212254.
\item K.Bardakci, Nucl.Phys. {\bf B 667} (2004) 354,
hep-th/0308197.
\item K.Bardakci, Nucl.Phys. {\bf B 698} (2004) 202,
hep-th/0404076.
\item K.Bardakci, Nucl.Phys. {\bf B 715} (2005) 141,
hep-th/0501107.
\item K.Bardakci, Nucl.Phys. {\bf B 746} (2006) 136,
hep-th/0602131.
\item  For some alternative approaches on putting field theory on
the world sheet, see O.Aharony, Z.Komargodski, S.S.Razamat,
JHEP {\bf 005} (2006) 016, hep-th/0602226, J.R.David, R.Gopakumar,
From Spacetime to Worldsheet: Four Point Correlators, hep-th/0606078,
A.Clark, A.Karch, P.Kovtun, D.Yamada, Phys.Rev. {\bf D 68} (2003)
066011, hep-th/0304107.
\item For an investigation of renormalization and Lorentz invariance
in the light cone formulation, see C.B.Thorn, Nucl.Phys. {\bf B 699}
(2004) 427, hep-th/0405018.
\item G.'t Hooft, Nucl.Phys. {\bf B 72} (1974) 461.
\item For a review of the large N method, see M.Moshe,
J.Zinn-Justin, Phys.Rep.{\bf 385} (2003) 69, hep-th/0306133.
\end{enumerate}

\end{document}